\documentclass[ukenglish]{nik}

\usepackage{graphicx}
\usepackage{subcaption}
\usepackage{amsmath}
\usepackage{latexsym} 
\usepackage{bm}
\usepackage{hyperref}
\usepackage[normalem]{ulem} 
\usepackage{amssymb,psfrag,pslatex,nicefrac,relsize}
\usepackage[table,xcdraw]{xcolor}
\usepackage{adjustbox}
\usepackage{multirow}
\usepackage{array}
\usepackage{booktabs}
\usepackage{listings}

\usepackage{mathptm} 

\usepackage[latin1]{inputenc}
 
\usepackage[T1]{fontenc} 

\usepackage{textcomp} 

\usepackage{type1cm} 

\title{An Evaluation of Score Level Fusion Approaches for Fingerprint and Finger-vein Biometrics}
\author{
        Kamer Vishi and Vasileios Mavroeidis\\ 
        \large Department of Informatics, SecurityLab \\ 
       \large University of Oslo (UiO), Norway\\ 
       \large \textit{{\{kamerv, vasileim\}@ifi.uio.no}}  
}

\begin{document}
\maketitle
\begin{abstract}

Biometric systems have to address many requirements, such as large population coverage, demographic diversity, varied deployment environment, as well as practical aspects like performance and spoofing attacks. Traditional unimodal biometric systems do not fully meet the aforementioned requirements making them vulnerable and susceptible to different types of attacks. In response to that, modern biometric systems combine multiple biometric modalities at different fusion levels. The fused score is decisive to classify an unknown user as a genuine or impostor. In this paper, we evaluate combinations of score normalization and fusion techniques using two modalities (fingerprint and finger-vein) with the goal of identifying which one achieves better improvement rate over traditional unimodal biometric systems. The individual scores obtained from finger-veins and fingerprints are combined at score level using three score normalization techniques (min-max, z-score, hyperbolic tangent) and four score fusion approaches (minimum score, maximum score, simple sum, user weighting). The experimental results proved that the combination of hyperbolic tangent score normalization technique with the simple sum fusion approach achieve the best improvement rate of 99.98\%.

\end{abstract}

\section{Introduction}
\label{sec:introduction}

The most important practical consideration for a biometric system is its accuracy. The combination of multiple different samples or multiple biometric modalities is a natural approach for improving the performance of a biometric system. This approach depends on specific methods for data fusion during the analytical process. Biometric fusion can be performed at different levels such as sensor level, feature (template) level, score level and decision level. Currently, the most popular and suitable way of biometric fusion is at score level since all commercially available biometric sensors and feature extractors do not provide access to their feature extraction algorithms. 

Some of the limitations of the aforementioned unimodal biometric systems are overcome by combining multiple biometric modalities. These systems are known as multimodal biometric systems and are more secure due to the integration of multiple independent pieces of evidence \cite{906041}. For instance, the chance of getting a valuable biometric system increases with the number of involved biometric traits. In addition, the integration of multiple traits increase security since it is more difficult to spoof multiple biometric modalities of real users \cite{Bhanu2011}. Advantages of multimodal biometric systems over unimodal systems have been discussed by Ross et al. \cite{Jain2004Best}. 

In this paper, we present an evaluation of normalization and fusion techniques using finger-vein and fingerprint biometrics and their potential application as biometric identifiers. The individual scores obtained from finger-vein and fingerprints are combined at score level using three score normalization techniques (Min-Max, Z-Score, Hyperbolic Tangent) and four score fusion approaches (Minimum Score, Maximum Score Simple Sum and User Weighting). The fused-score classifies an unknown user as genuine or impostor.

The rest of the paper is organized as follows: Section~\ref{sec:related_work} gives an overview of related work, Section~\ref{sec:scorenormfuse} describes the normalization and fusion techniques, Section~\ref{sec:experiments} presents the experiments conducted and the data analysis, Section~\ref{sec:results} presents the experimental results, and lastly, Section~\ref{sec:conclusion} rounds off the paper with a discussion and indication of future work.

\section{Related Work}
\label{sec:related_work}
 Different score fusion approaches have been proposed for fusing scores received from different biometric modalities.

Raghavendra et al. \cite{Raghavendra2009} proposed a multimodal biometric score level fusion scheme using Gaussian Mixture Model and Monte Carlo Method. The authors fused face, speech and palmprint modalities and showed that their method gives higher performance than other fusion schemes, such as Simple sum rule, Weighted sum rule, Fishers Linear Discriminate Analysis (FLD), and Likelihood Ratio (LR) methods.

Derawi et al. \cite{Derawi2010} fused gait and fingerprint traits for user authentication on mobile devices. The authors used four different methods to normalize the scores (min-max, z-score, median absolute deviation, tangent hyperbolic) and four fusion approaches (simple sum, user-weighting, maximum score and minimum score). The fusion results of fingerprint and gait recognition showed an improved performance over other methods and brought as a step closer to the advancement of user authentication on mobile devices.

Peng et al. \cite{Peng2014} fused finger vein, fingerprint, finger shape and finger knuckle print of a single human finger based on triangular norm (t-norm). The results showed that score level fusion using triangular norm obtains a larger distance between genuine and imposter score distribution, as well as lower error rates.

Conti et al. \cite{Conti2007Fuzzy} and Lau et al. \cite{Lau2004} proposed a multi-instance fusion approach. Their approach is based on fusing two comparison scores using fuzzy logic rules from two different fingerprints. Experimental results showed an improvement of 6.7\% using the comparison score level fusion rather than a unimodal authentication system. 

Kisku et al. \cite{Kisku2009} proposed a multimodal biometric system using face and ear biometrics. They used Gaussian Mixture Model (GMM) with belief fusion for the estimated scores characterized by Gabor responses, and the proposed fusion is accomplished by Dempster-Shafer (DS) decision theory. It has been proved that DS provides increased accuracy and significant improvements over the existing classical fusion rules.

\section{Score Level Fusion}\label{sec:scorenormfuse}
\begin{figure}[h]
\centerline{\includegraphics[width=1\textwidth]{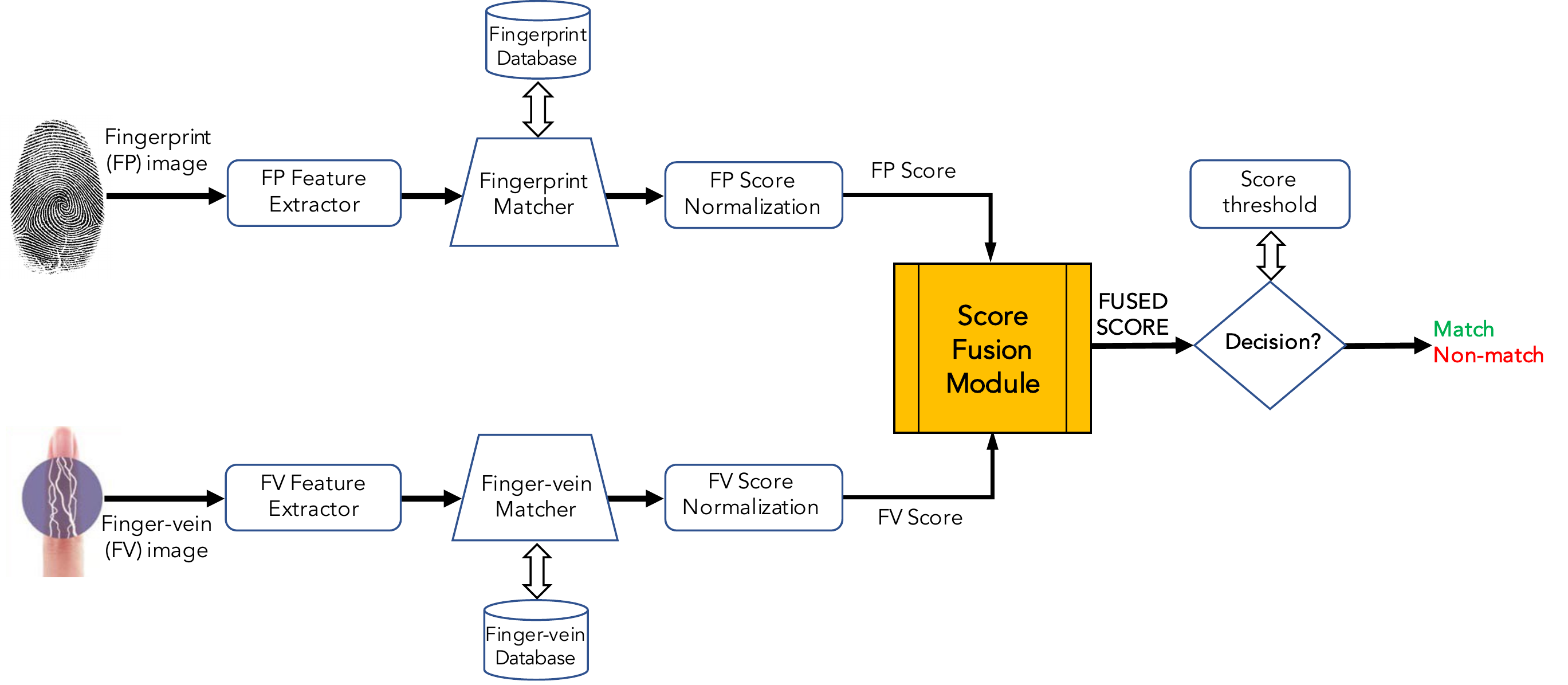}}
\caption{Score Level Fusion Diagram for Fingerprint and Finger-vein.}
\label{fig:score_fusion}
\end{figure}

Most of the multimodal biometric systems integrate data at score level due to the strong trade-off between the ease in combining the data and better information content. Besides, it is a relatively straightforward way to combine scores generated by different comparators (matchers)\footnote{Note that the term \textit{"matching"} as a synonym for \textit{"comparison"} has been deprecated in the ISO SC37 Harmonized Biometric Vocabulary \cite{ISO_sc37}.}. Therefore, score level fusion is the preferred approach for integrating biometric data. Figure \ref{fig:score_fusion} illustrates this process. 

At score level, fused scores of each modality are combined with a variety of techniques to produce a new score that is compared with the threshold in the decision module. There are two key approaches in use today for consolidating comparison scores; classification and combination. In \textit{classification} we can construct a  vector with individual scores that are classified into accept or reject classes. A classification approach might use a decision tree, Support Vector Machine (SVM), or Linear Discriminate Analysis (LDA) to classify the vector as an impostor or genuine. In \textit{combination} approach, individual scores are combined to generate a single scalar score to render the final decision \cite{Vishi:2017:NAM:3109761.3158409}. The combination approach for consolidating scores has compiled a superior performance record versus the other levels.

Score level fusion has two main steps. The first step is \textit{score normalization}, where calculated scores are created by a certain comparator \begin{math} S_i \end{math} and are mapped onto a new score scale or domain \begin{math} S_{i}^{'} \end{math}.
For instance, if a comparator X produces scores on a domain $[1, 100]$ and a comparator Y generates scores on a domain $[1, 2500]$ these scores need to be normalized and mapped at a common scale.
The second step of fusion at score level is \textit{fusion}. There are many score fusion techniques, however, in this paper, we have used only some of the fusion techniques that are recommended by ISO standards -  \textit{"ISO/IEC TR 24722:2007 - Multimodal and other Multibiometric Fusion"}\cite{iso_fusion} and are described in Section \ref{sub:score_fusion_tech}. 

\subsection{Score Normalization}\label{sub:score_normalization}
Score normalization processes have been researched extensively \cite{Vishi:2017:NAM:3109761.3158409}. In this section, we describe some fundamental normalization concepts used in biometrics to facilitate easier interpretation of our paper.
The score normalization process is performed to transform the comparator's parameters and data types into a common domain. Commonly, score normalization techniques are evaluated based on their robustness and efficiency.

The most used score normalization techniques that are also used and evaluated in this paper are Min-Max (MM), Z-Score (ZS), and Hyperbolic Tangent (TanH). These methods are discussed below.

\begin{table}[ht]
\centering 
\caption{Symbols Used for Score Normalization Expressions.}
\begin{tabular}{|l|c|c|c|}
    \hline
    {\bf Statistical Measures} & {\bf Genuine Distribution} & {\bf Impostor Distribution} & {\bf Both} \\ [1.5ex]
    \hline
    Minimum Score & $S_{Min}^{G}$         & $S_{Min}^{I}$     & $S_{Min}^{B}$        \\ [1.5ex]
    \hline
    Maximum Score & $S_{Max}^{G}$     & $S_{Max}^{I}$     & $S_{Max}^{B}$             \\ [1.5ex]
    \hline
    Mean          & $S_{Mean}^{G}$     & $S_{Mean}^{I}$   & $S_{Mean}^{B}$            \\ [1.5ex]
    \hline
    Score Standard Deviation   & $S_{SD}^{G}$ & $S_{SD}^{I}$   & $S_{SD}^{B}$         \\ [1.5ex]
    \hline
\end{tabular}
\label{table:symbolsmulti} 
\end{table}
\begin{description}
\item[Min-Max Normalization (MM)] performs a linear transformation of the original data. This is one of the simplest normalization techniques and is appropriate when the limits of the produced scores are known. MM is considered efficient and provides adequate performance; however, it may not yield completely accurate results if the data contain outliers. MM maps raw scores to the [0,1] range and given scores S$_{Max}^B$ and S$_{Min}^B$ designates the endpoints of the score range.

\begin{equation}
S' = \frac{S-S_{Min}^{B}}{S_{Max}^{B} - S_{Min}^{B}}
\end{equation}

\item[Z-Score Normalization (ZS)] is one of the most commonly used normalization techniques. It uses an arithmetic mean and a standard deviation to normalize the data; therefore, a priori knowledge regarding the average score and score variances of the comparator is needed. ZS is considered efficient and tends to work exceptionally well if the scores of each modality follow a Gaussian distribution; however, this technique may not achieve similar accuracy if the data used contain outliers (the mean and standard deviation are sensitive to outliers). ZS normalization transforms the scores to a normal distribution with an arithmetic mean S$_{Mean}^I$ of 0 and a standard deviation S$_{SD}^I$ of 1.
\begin{equation}
S' = \frac{S-S_{Mean}^{I}}{S_{SD}^{I}} 
\end{equation}

\item[Hyperbolic Tangent Normalization (TanH)] is efficient (when the parameters are selected carefully), provides adequate performance, and is very robust in handling outliers. TanH maps the raw scores to the (0,1) range, where S$_{Mean}^G$ and S$_{SD}^B$ are the mean and standard deviation estimation of the score distribution, respectively.
\begin{equation}
S' = 0.5 \cdot tanh \cdot \frac{0.01 (S-S_{Mean}^{G})}{S_{SD}^{B}} + 1
\end{equation}

\end{description}

\begin{figure}[h!] 
\centering
\includegraphics[width=1.0\textwidth]{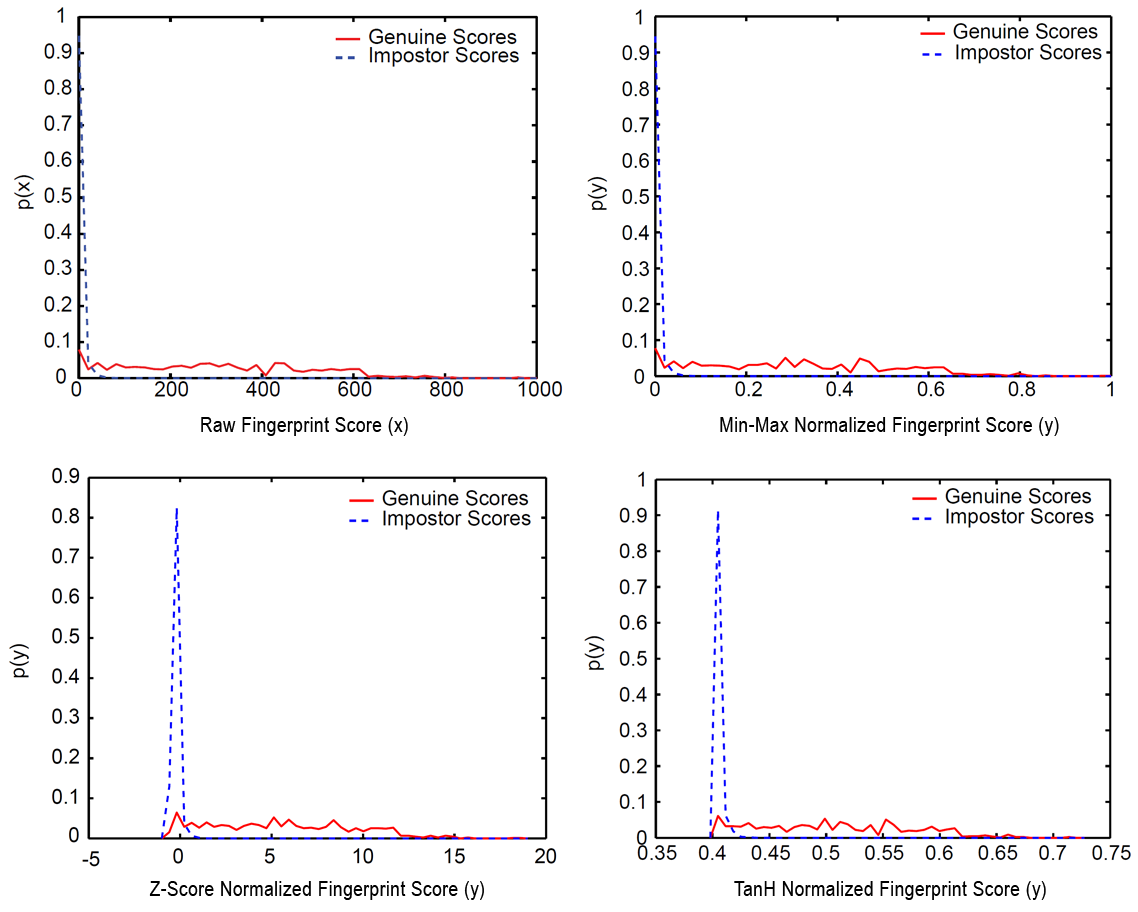}
\caption[Distributions of Genuine and Impostor Comparison Scores.]{Distributions of Genuine and Impostor Comparison Scores.}
\label{fig:score_normal1}
\end{figure}

Figure \ref{fig:score_normal1}, illustrates distribution graphs for the aforementioned fingerprint score normalization techniques; where x represents raw scores before normalization and y represents normalized scores.

\subsection{Score Fusion Techniques}\label{sub:score_fusion_tech}
In general, score fusion techniques fall under two categories: classification and combination.
\textit{Classification} techniques formulate the problem by dividing the decision space into two classes: genuine and impostor. The reliability and effectiveness of this method are dependent on the large amount and quality of the input data that are used to train the classifier. In addition, the comparison scores do not need to be homogeneous; hence the normalization step is not required. Some of the classification methods that have been researched are neural networks, nearest neighborhood algorithms, and tree-based classifiers.

\textit{Combination} is the most common and effective technique for combining biometric scores. This technique combines scores from multiple comparators and generates a single score. Combination requires score normalization before fusing the scores. This paper analyses the following score level fusion techniques: Maximum Score, Minimum Score, Simple Sum, and Weighted Sum.

\begin{description}
\item[Maximum Score:] the max rule estimates the mean of the posteriori probabilities by the maximum value.
\begin{equation}
max (i=1\: to\: N)\: S_i'
\end{equation}
\item[Minimum Score:] the min rule sets the minimum value of posteriori probabilities.
\begin{equation}
min (i=1\: to\: N)\: S_i'
\end{equation}
\item[Simple Sum:] it is a weighted average of the raw scores. Comparison scores are summed without the benefit of normalization routines. It simplistically assumes that the raw scores supplied by the biometric methods used have a comparable scale, distribution, and strength. It can be used in ambiguous classifications resulting from high-level noise.
\begin{equation}
\sum (i=1\: to\: N)\: S_i'
\end{equation}
\item[Weighted Sum:] this method computes the combined score as a weighted sum of the comparison scores. The motivation behind the idea of user-specific weights for computing a weighted sum of scores is that some biometric traits cannot be reliably obtained from some people (e.g., individuals with faint fingerprints). Assigning a lower weight to a fingerprint score and a higher weight to a finger-vein score reduces the probability of a false rejection.
\begin{equation}
\sum (i=1\: to\: N)\: W_i^* \cdot S_i'
\end{equation}
\end{description}
 When it is implemented correctly, score level fusion can improve accuracy, thwart fraudsters, and increase usability. On the other hand, in case it is implemented incorrectly, a multibiometric system might experience performance degradation in comparison to a unimodal system. Furthermore, multimodal biometric systems have some disadvantages. They could potentially have a higher cost of ownership, they could increase user inconvenience, they could decrease user acceptance, and also exacerbate privacy issues.

\section{Experiment Setup}
In this section we present details regarding the database, datasets and modifications needed to conduct our experiments.
\label{sec:experiments}
\subsection{Database}
Fingerprint and finger-vein experiments in this paper are conducted over four different datasets collected during the summer of 2010 by the Machine Learning and Applications (MLA) Group at Shandong University, in China \cite{Yin2011}.

\begin{figure}[h!] 
\centering
\includegraphics[width=0.6\textwidth]{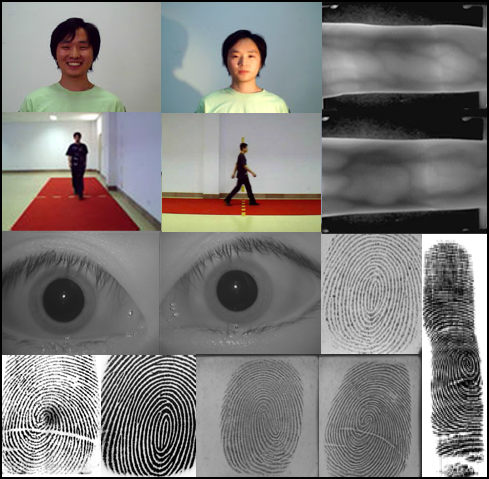}
\caption[SDUMLA-HMT Database samples]{Sample Images of Face, Finger-vein, Gait, Iris and Fingerprint from SDUMLA-HMT Database \cite{Yin2011}.}
\label{fig:sdumla_db}
\end{figure}

The MLA group named this database \textit{''SDUMLA- HMT: A Multimodal Biometric Database''}. The database includes 106 subjects; 61 males and 45 females between 17 and 31 years old \cite{Yin2011}. The database consists of face images captured from seven different view angles, finger vein images of six fingers, gait videos from six view angles, finger-vein images from a finger-vein sensor and fingerprint images captured by five different sensors. Figure \ref{fig:sdumla_db}, shows some sample images from SDUMLA-HMT database.

 In addition, modifications were needed over the acquired data to conduct our experiment. In particular,  we have reduced the number of participants, fingers, and impressions as follows: we use two index fingers of both hands out of 6 fingers provided in the dataset, five impressions for each fingerprint out of eight that were captured, and we reduced the number of participants from 106 to 100. These modifications were demanded because in the provided database there were available only finger-vein images of the index fingers. Consequently, we have \begin{math}
2(fingers)x100(subjects)x5(attempts)=1000\: images
\end{math} per dataset. In other words, we used 2000 fingerprint images from two datasets (DS2 and DS3) out of a total of 25,440 fingerprint images.
Furthermore, we decided to follow the naming convention recommended in ISO 19794-2 \cite{iso_finger}. To do that we requested the original finger-code positions from the MLA Group and we renamed them accordingly.

\begin{figure}[h!] 
\centering
\includegraphics[width=1.0\textwidth]{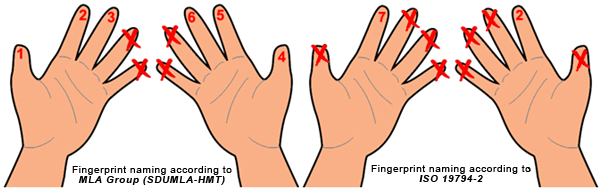}
\caption[Illustration of Finger Position Codes.]{Illustration of Finger-code Positions (names).}
\label{fig:finger_names}
\end{figure}

An illustration of finger-code positions (names) from the MLA Group and ISO 19794-2 is given in Figure \ref{fig:finger_names}. 

\subsection{Fusion Experiments}\label{sec:fusion_experiments}
Based on a quality assessment conducted over the database, the best datasets are named with suffix 1, while the worst datasets are named with suffix 2.
\begin{itemize}
\item \textbf{Fingerprint Datasets:}
\begin{itemize}
    \item Fingerprint, best quality dataset (SDUMLA-HMT DS2) is named as FP\_DS1,
    \item Fingerprint, worst quality dataset (SDUMLA-HMT DS3) is named as FP\_DS2.
\end{itemize}
\item \textbf{Finger-vein Datasets:}
\begin{itemize}
    \item Finger-vein, best quality dataset (SDUMLA-HMT Finger-vein-Lamp) is named as FV\_DS1,
    \item Finger-vein, worst quality dataset (SDUMLA-HMT Finger-vein) is named as FV\_DS2.
\end{itemize}
\end{itemize}
We have defined four fusion scenarios; fusion with the best and the worst dataset as follows:
\begin{enumerate}
\item Fusion of  \textit{FP\_DS1} and \textit{FV\_DS1}
\item Fusion of  \textit{FP\_DS1} and \textit{FV\_DS2}
\item Fusion of  \textit{FP\_DS2} and \textit{FV\_DS1}
\item Fusion of  \textit{FP\_DS2} and \textit{FV\_DS2}
\end{enumerate}

\section{Experimental Results}
\label{sec:results}
We have retrieved both low and high Equal Error Rates (EERs) from unimodal biometric systems. In table \ref{tab:comparison_of_methods}, we can see that TanH gives better performance than MinMax and Z-Score. Additionally, Simple Sum gives better results than Maximum Score, Minimum Score and User Weighted Sum. Specifically, Simple Sum yields 99.98 \% average verification accuracy, making it the most stable fusion framework.

\begin{table}[!ht]
\centering
\caption{Sample Comparison Results for Different Normalization and Fusion Techniques.}
\label{tab:comparison_of_methods}
\begin{tabular}{|l|c|c|c|}
    \hline
    \multicolumn{4}{|c|}{\cellcolor[HTML]{FFCB2F}{\color[HTML]{FE0000} \textbf{EER (\%)}}}                                                                                                   \\ \hline
    \multicolumn{1}{|c|}{\cellcolor[HTML]{BBDAFF}}                                                     & \multicolumn{3}{c|}{\cellcolor[HTML]{BBDAFF}\textit{\textbf{Normalization Technique}}} \\ \cline{2-4} 
    \multicolumn{1}{|c|}{\multirow{-2}{*}{\cellcolor[HTML]{BBDAFF}\textit{\textbf{Fusion Technique}}}} & MinMax(MM)                  & Z-Score(ZS)                 & TanH                 \\ \hline
    Minimum Score (MinS)                                                                               & 3.81836\%                     & 7.32851\%                     & 1.01881\%                    \\ \hline
    Maximum Score (MaxS)                                                                               & 0.11591\%                     & 0.11206\%                     & 0.07837\%                    \\ \hline
    User Weighting (UW)                                                                                    & 0.08949\%                    & 0.15935\%                     & 0.01763\%                   \\ \hline\textbf{Simple Sum (SS)}                                                                              & 0.08281\%                     & 0.10955\%                     & \textbf{0.00010\%}                   \\ \hline
\end{tabular}
\end{table}
\subsection{Fusion Scenarios}

In the following sections we summarize the results of four fusion scenarios (fusion with the best and the worst dataset).
\subsubsection{ \textbf{Scenario \#1:\textit{Fusion of FP\_DS1 and  FV\_DS1}}}
This case shows the fusion performance of finger-vein dataset \textit{FV\_DS1} and fingerprint dataset \textit{FP\_DS1} using Hyperbolic Tangent estimators (TanH) normalization and Simple Sum (SS). The results show that the fingerprints dataset attains an EER of 0.86\%, the finger-veins dataset an EER of 0.71\%, and their fusion a low EER of \textit{0.00010\%}.

\subsubsection{\textbf{Scenario \#2: \textit{Fusion of FP\_DS1 and FV\_DS2}}}
This scenario shows the fusion performance of the fingerprint dataset \textit{FP\_DS1} and finger-vein dataset \textit{FV\_DS2} using Hyperbolic Tangent (TanH) and Maximum Score (MaxS). 
The results show that the fingerprints dataset attains an EER of 0.86\%, the finger-veins dataset an EER of 7.35\%, and their fusion a low EER of \textit{0.0320\%}.


\subsubsection{\textbf{Scenario \#3: \textit{Fusion of FP\_DS2 and FV\_DS1}}}
This case shows the fusion performance of the fingerprint \textit{FP\_DS2} and the finger-vein datasets \textit{FV\_DS1} using MinMax normalization (MM) and Maximum Score (MaxS) fusion rule. The results show that the fingerprints dataset attains an EER of 1.01\%, the finger-veins dataset an EER of 0.71\%, and their fusion a low EER of \textit{0.00015\%}.


\subsubsection{\textbf{Scenario \#4: \textit{Fusion of FP\_DS2 and FV\_DS2}}}

\begin{figure}[h!]
\centering
\includegraphics[width=1\textwidth]{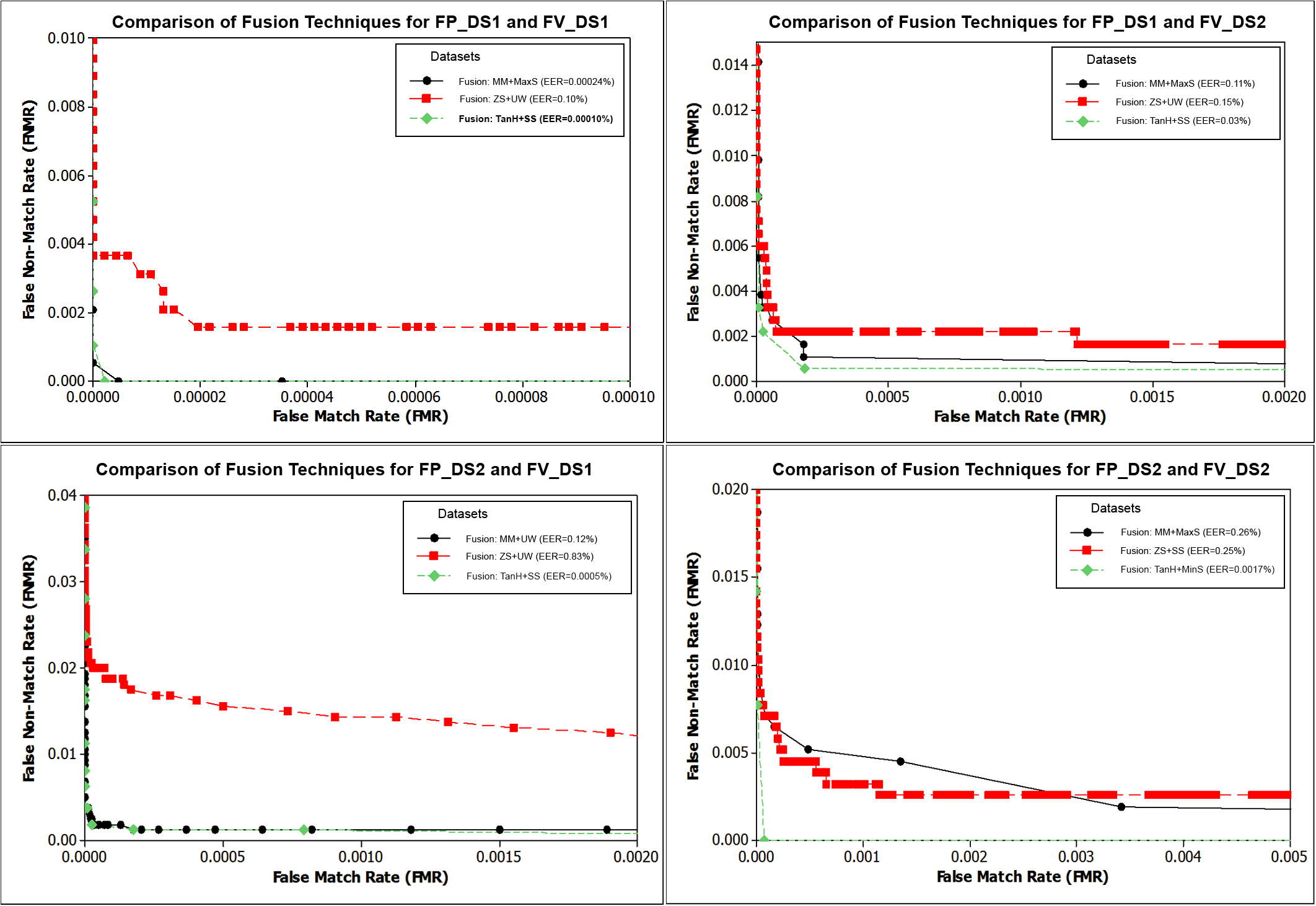}
\caption{Performance of Fingerprint and Finger-vein Using TanH Score Normalization and Simple Sum Score Fusion}
\label{fig:fusion_comparison}
\end{figure}

This case shows fusion performance of the fingerprint  \textit{FP\_DS2} and finger-vein datasets \textit{FV\_DS2} using Hyperbolic Tangent estimators (TanH) normalization and Maximum Score fusion rule. The results show that the fingerprints dataset attains an EER of 1.01\%, the finger-veins dataset an EER of 7.35\%, and their fusion a low EER of \textit{0.0038\%}.\\\\
Figure \ref{fig:fusion_comparison}, presents four graphs for different normalization and fusion techniques per fusion scenario (we get an insight of the increased performance when TanH and SS is used, in relation to other normalization and fusion techniques).


To summarize, the results indicate that fusion of finger-vein and fingerprint improves the performance of biometric systems. We can define \textit{improvement} as the percentage difference between the minimum EER of fingerprint or finger-vein (we choose the lowest value between the two) minus the EER of their fusion and (divided by) the minimum EER of fingerprint or finger-vein (we choose the lowest value between the two):
\begin{eqnarray}
Improvement &=& \frac{EER_{min(print | vein)} - EER_{fusion(print+vein)}}{EER_{min(print | vein)}} \cdot 100 
\label{eq:improvementsEER}
\end{eqnarray}

Table \ref{tab:fusion_improvements} summarizes our results. The values in percentages indicate EER. 

\begin{table}[h!]
\centering 
\begin{adjustbox}{width=1\textwidth}
    \begin{tabular}{c|c|c|c} 
        \hline\hline 
        \textbf{Fingerprint alone} & \textbf{Finger-vein alone}  & \textbf{Fusion$_{(print + vein)}$}  & \textbf{Improvement (method)} \\ [0.5ex] 
        \hline %
        0.86 \%    & 0.71 \% & 0.0001 \%  & 99.98 \% (TanH+SS) \\
        \hline %
        0.86 \%    & 7.35 \%  & 3.81 \%  & 48.10 \% (MM+MinS)\\
        \hline %
        1.01 \%    & 0.71 \% & 0.1872 \%  & 79.50 \% (TanH+UW)\\
        \hline %
        1.01 \%    & 7.35 \%  & 0.0038 \%  & 89.62 \% (ZS+SS) \\ 
        \hline 
    \end{tabular}
\end{adjustbox}
\caption[Multimodal Fusion Improvements of Fingerprint and Finger-vein Recognition.]{Multimodal Fusion Improvements of Fingerprint and Finger-vein Recognition.}
\label{tab:fusion_improvements}
\end{table}
\section{Conclusion}
\label{sec:conclusion}
Multibiometric systems are the new frontier, and this is evident from the number of ongoing efforts in the research and commercial domains. These systems address some of the deficiencies mentioned above of the unimodal biometric systems. Theoretically, they present a huge potential, but research and commercial systems have yet to reflect this promise. Furthermore, the lack of standards indicates that more work is required before it reaches an acceptable level of maturity.

Experimental results show that in most cases fused performance (fingerprint and finger-vein) was significantly improved compared to unimodal biometric. It is worth noting that the best fusion performance is achieved by the combination of Hyperbolic Tangent (TanH) score normalization technique and Simple Sum (SS) method for fusion, yielding an EER of 0.00010\%. In other words, a multimodal biometric system with fingerprint and finger-vein would perform 99.98\% better than a unimodal biometric system of fingerprint or finger-vein. In addition, it has been observed that both MinMax and Z-Score methods are sensitive to outliers. On the other hand, TanH score normalization is both \textit{robust} and \textit{efficient}.

Nevertheless, our future work should investigate fusion approaches using fingerprint and finger-vein biometrics at feature extraction level or template level, to secure biometric templates and enhance privacy by hiding the meaning of extracted features (points) from finger-vein and fingerprint in the stored template.

\section*{Acknowledgement}
This research is supported by the Research Council of Norway under the Grant No.: IKTPLUSS 248030/O70 and 247648 for SWAN and Oslo Analytics projects, respectively. This research is also part of the SecurityLab of University of Oslo.

%
\newpage

\bibliographystyle{unsrt} 
\bibliography{nisk-references}

\end{document}